\documentclass[lettersize,journal]{IEEEtran}
\usepackage{amsmath,amsfonts}
\usepackage{algorithmic}
\usepackage{algorithm}
\usepackage{array}
\usepackage[caption=false,font=normalsize,labelfont=sf,textfont=sf]{subfig}
\usepackage{textcomp}
\usepackage{stfloats}
\usepackage{url}
\usepackage{verbatim}
\usepackage{graphicx}
\usepackage{cite}
\usepackage{bm}
\usepackage{amssymb}
\usepackage{subfloat}
\usepackage{xcolor}
\usepackage{xpatch}

\definecolor{ChangeColor}{rgb}{1,0,0}
\def\BibTeX{{\rm B\kern-.05em{\sc i\kern-.025em b}\kern-.08em
   T\kern-.1667em\lower.7ex\hbox{E}\kern-.125emX}}
\begin{document}

\title{Single-BS Simultaneous Environment Sensing and UE Localization without LoS Path by Exploiting Near-Field Scatterers}

\author{Zhiwen~Zhou, Zhiqiang~Xiao,~\IEEEmembership{Graduate Student Member,~IEEE},
        and
        Yong~Zeng,~\IEEEmembership{Senior Member,~IEEE}\vspace{-2em}
\thanks{This work was supported by the Natural Science Foundation for Distinguished Young Scholars of Jiangsu Province with number BK20240070. The authors are with the National Mobile Communications Research Laboratory, Southeast University, Nanjing 210096, China. Zhiqiang Xiao and Yong Zeng are also with the Purple Mountain Laboratories, Nanjing 211111, China (e-mail: {zhiwen\_zhou, zhiqiang\_xiao, yong\_zeng}@seu.edu.cn).}
}

\maketitle

\begin{abstract}
As the mobile communication network evolves over the past few decades, localizing user equipment (UE) has become an important network service. While localization in line-of-sight (LoS) scenarios has reached a level of maturity, it is known that in far-field scenarios without a LoS path nor any prior information about the scatterers, accurately localizing the UE is impossible. In this letter, we show that this becomes possible if there are scatterers in the near-field region of the base station (BS) antenna arrays. Specifically, by exploiting the additional distance sensing capability of extremely large-scale antenna arrays (XL-arrays) provided by near-field effects, we propose a novel method that simultaneously performs environment sensing and non-line-of-sight (NLoS) UE localization using one single BS. In the proposed method, the BS leverages the near-field characteristics of XL-arrays to directly estimate the locations of the near-field scatterers with array signal processing, which then serves as virtual anchors for UE localization. Then, the propagation delay for each path is estimated and the position of the UE is obtained based on the positions of scatterers and the path delays. Simulation results demonstrate that the proposed method achieves superior accuracy and robustness with similar complexity compared with benchmark methods.
\end{abstract}


\IEEEpeerreviewmaketitle
\section{Introduction}
User equipment (UE) localization is a crucial functionality in modern mobile communication networks, which has found a wide range of applications such as location-based services, emergency response coordination, transportation logistics, and indoor navigation\cite{xiao2022overview}\cite{localization_application2}. Besides, UE localization can also enhance communication capabilities, such as optimizing handover management\cite{malm2018user} and enabling location and environment-aware communications by techniques like channel knowledge map (CKM) \cite{zeng2021toward}.

Wireless localization methods typically rely on three types of measurements: received signal strength (RSS), time of arrival (ToA), and angle of arrival (AoA). While RSS-based methods are usually easy to implement, they typically have inferior performance than ToA- and AoA-based methods.  However, in practical environments, the presence of NLoS propagation paths poses challenges to UE localization. For ToA-based methods, the accuracy of UE distance estimation may be compromised in the presence of NLoS paths\cite{he2012modeling}. Nonetheless, as long as the LoS path exists, UE distance estimation remains feasible by recognizing that the LoS path corresponds to the smallest delay. However, methods based solely on AoA measurements become ineffective since multiple AoAs are measured from a single UE. Additional measurements, such as the ToA or RSS of each path, are necessary for differentiating the LoS path from the NLoS paths.

The biggest challenge occurs when no LoS path exists between the UE and the base station (BS). It has been theoretically shown that in far-field scenarios, NLoS signals do not contribute to the equivalent Fisher information matrix (EFIM) for the UE's position and thus can not improve localization accuracy if no prior information about the scatterers is available \cite{shen2010fundamental}.  As a result, without a LoS path nor any prior information of the scatterers, accurate UE localization is impossible in far-field scenarios. Numerous efforts have been made to address localization in purely NLoS environments by assuming some prior information about the scatterers. These methods can be categorized into two groups. The first type employs statistical methods, assuming  some known probability distribution of scatterers to obtain statistically optimal estimates of the UE location\cite{al2002ml,liu2003toa}. The second type aims to estimate the UE location using geometric methods, specifically by leveraging the geometric relationships between the UE and the scatterers for localization. Reconfigurable intelligent surfaces (RISs) have also been considered to tackle this problem in both near-field and far-field scenarios\cite{cao2024unified}. Nevertheless, these methods have limited applications due to their reliance on RISs or known scatterer locations. 

With the introduction of massive antenna arrays, estimation of scatterer locations using one single BS becomes possible by exploiting the near-field effects\cite{wang2024cramer}. Specifically, the array response vector of a massive array is dependent on both the AoA and the distance of the signal source. Leveraging on this property, \cite{tensormethod} achieves near-field localization of UE in NLoS scenarios. However, the tensor-based method proposed in \cite{tensormethod} requires each path's parameters, i.e., delay, Doppler shift and scatterer position to be different for the canonical polyadic (CP) decomposition uniqueness conditions to hold. If any of the parameters are too close, e.g., two paths have close delays, Doppler shifts or two scatterers are physically close, thus having similar array response vectors, the paths may not be properly resolved due to rank deficiencies.

In this letter, by exploiting the channel spatial sparsity and high spatial resolution of extremely large-scale antenna arrays (XL-arrays), we propose a novel scheme to solve the challenging problem of single-BS simultaneous environment sensing and UE localization in the absence of LoS path. This is possible since the locations of near-field scatterers can be directly estimated using array signal processing techniques. As such, these near-field scatterers can function as virtual anchors for UE localization. 
Furthermore, as shown in Section \ref{ueloc}, our scheme enables the estimation of clock difference between BS and UE using four or more near-field scatterers. This capability eliminates the necessity for perfect synchronization between BS and UE. The estimated clock difference can also be exploited to improve BS-UE clock synchronization accuracy, solving the synchronization headache in bi-static sensing. 

\section{System Model}
\label{sysmodel}
\begin{figure}[!htbp]
\vspace*{-8pt}
\centering
\includegraphics[width=0.48\textwidth]{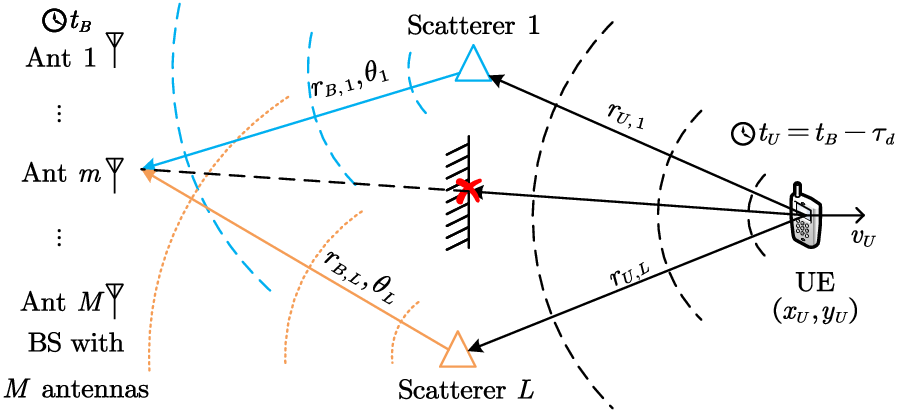}
\hfil
\vspace*{-6pt}
\caption{A single BS localization system with clock asynchronism and without LoS path.}
\label{scene}
\vspace*{-6pt}
\end{figure}
As shown in Fig. \ref{scene}, we consider a single BS localization system where the UE has one antenna and the BS is equipped with an $M$-element extremely large-scale uniform linear array (XL-ULA). Let $L$ denote the number of scatterers. $r_{U,l}$ and $r_{B,l}$ are the distance from scatterer $l$ to the UE and the reference antenna element of the BS XL-ULA, respectively, and $\theta_l\in [-\frac{\pi}{2},\frac{\pi}{2})$ is the AoA of the $l$th scatterer with respect to the reference antenna element of the BS. $\tau_l=(r_{B,l}+r_{U,l})/c$ is the propagation delay of the $l$th multi-path and $\tau_d$ is the clock difference between the UE and the BS. The UE is moving with velocity $\boldsymbol{v}_U\in\mathbb{R}^{2 \times 1}  $. Note that the proposed method still works well when $\boldsymbol{v}_U=\boldsymbol{0}$. Our goal is to achieve environment sensing and NLoS UE localization, i.e., to estimate the locations $\{(r_{B,l},\theta_l)\}_{l=1}^{L}$ of the scatterers and the location $(x_U,y_U)$ of the NLoS UE.

In order to estimate the UE location $(x_U,y_U)$, the following equations needs to be solved,
\begin{equation}
\begin{aligned}
&\left\{ \begin{array}{l}
	(r_{B,l}\cos \theta _l-x_U)^2+(r_{B,l}\sin \theta _l-y_U)^2=r_{U,l}^{2}\\
	r_{U,l}+r_{B,l}=c\tau _l\\

\end{array}\right.\\
&\forall l=1,\ldots,L.
\end{aligned}
\end{equation}
In far-field scenarios with traditional smaller-scale arrays, assuming perfect synchronization of clocks, i.e., $\tau_d=0$, the variables $\{\theta_l\}_{l=1}^L$ and $\{\tau_l\}_{l=1}^L$ can be estimated from the received signal. However, other variables remain unknown, resulting in an under-determined system with $2L+2$ unknowns and only $2L$ equations. On the other hand, in near-field scenarios with XL-arrays, provided that $L\ge3$, the additional estimation of $\{r_{B,l}\}_{l=1}^L$ through array signal processing leads to an over-determined system, enabling the estimation of UE's location.


The channel impulse response is written as
\begin{equation}
\abovedisplayshortskip=2pt
\belowdisplayshortskip=2pt
\abovedisplayskip=2pt
\belowdisplayskip=2pt
\begin{aligned}
\boldsymbol{h}\left( t, \tau \right) =\sum_{l=1}^L{\boldsymbol{h}_l}\delta \left( \tau-\tau_{s,l}\right)e^{j2\pi f_{D,l}t} ,
\end{aligned}
\label{SIMOchannel}
\end{equation}
where $\bm{h}_l \in \mathbb{C} ^{M\times 1}$ denotes the channel coefficient vector and $\tau_{s,l}=\tau_l+\tau_d$. $\boldsymbol{h}_l=\alpha _l\boldsymbol{a}\left( r_{B,l},\theta _l \right) $, with $\alpha _l$ denoting the complex-valued path gain. $f_{D,l}$ denotes Doppler shift of path $l$. We adopt the uniform spherical wave (USW) model \cite{nearfield} here, in which $\boldsymbol{a}\left( r_{B,l},\theta _l \right)$ is the near-field array response vector for the $l$th scatterer, given by
\begin{equation}
\abovedisplayshortskip=2pt
\belowdisplayshortskip=2pt
\abovedisplayskip=2pt
\belowdisplayskip=2pt
\begin{aligned}
\boldsymbol{a}\left( r_{B,l},\theta _l \right) =\left[ 1,e^{-j\phi_{l,2}},...,e^{-j\phi_{l,M}} \right] ^T,
\end{aligned}
\label{ArrayVector}
\end{equation}
where the reference element is the first antenna element and the phase difference between the $m$th element and the first element $\phi_{l,m}$ can be expressed as
\begin{equation}
\abovedisplayshortskip=2pt
\belowdisplayshortskip=2pt
\abovedisplayskip=2pt
\belowdisplayskip=2pt
\begin{aligned}
\phi_{l,m}\!=\!\frac{2\pi}{\lambda}{\left(\sqrt{r_{B,l}^2\!-\!2r_{B,l}(m\!-\!1)d\sin \theta_l\!+\!(m\!-\!1) ^2d^2}\!-\!r_{B,l}\right)}.
\end{aligned}
\label{Deltaphi}
\end{equation}
Note that the USW model used here is more general and applicable to both near- and far-field scenarios. For far-field scatterers, $R_{B,l}$ is larger than the Rayleigh distance. Assume that the UE transmits OFDM signal $s(t)$ with power $P$, i.e.,
\begin{equation}
\abovedisplayshortskip=2pt
\belowdisplayshortskip=2pt
\abovedisplayskip=2pt
\belowdisplayskip=2pt
\begin{aligned}
&s\left( t \right) = \sum_{\gamma=0}^{\varGamma-1}{\!}\sum_{n=0}^{N-1}{b_{n,\gamma}}e^{j2\pi n\Delta f\left( t\!-\!\gamma T_O\!-\!T_{CP} \right)}\mathrm{rect}\!\left( \!\frac{t\!-\!\gamma T_O}{T_O}\! \right) \!,
\end{aligned}
\label{Signal}
\end{equation}
where $\Delta f$ is the subcarrier spacing, $T_{CP}$ is the duration of the cyclic prefix (CP), $T_O$ is the duration of the OFDM symbol, $\{b_{n,\gamma}\}_{n=0,\gamma=0}^{n=N-1,\gamma=\varGamma-1}$ contains $N\varGamma$ independent and identically distributed (i.i.d) information-bearing symbols such as the quadrature amplitude modulated (QAM) symbols, with element ${b_{n,\gamma}}$ denoting the symbol on the $n$th subcarrier and $\gamma$th OFDM symbol with $\mathbb{E} \left\{ \left|b_{n,\gamma} \right|^2 \right\} =P/N$.  The received signal at the BS is
\begin{equation}
\abovedisplayshortskip=2pt
\belowdisplayshortskip=2pt
\abovedisplayskip=2pt
\belowdisplayskip=2pt
\begin{aligned}
\boldsymbol{y}\left( t \right) &=\int_{-\infty}^{\infty}{\boldsymbol{h}\left( t,\tau \right) s\left( t-\tau \right)}+\boldsymbol{n}\left( t \right) 
\\
&=\sum_{l=1}^L{\boldsymbol{h}_ls\left( t-\tau _{s,l} \right)}e^{j2\pi f_{D,l}t}+\boldsymbol{n}\left( t \right) ,
\end{aligned}
\label{ReceivedSignal}
\end{equation}
where  $\boldsymbol{n}\left( t \right)$ is the additive Gaussian noise with power $\sigma^2$. The BS samples the signal $\boldsymbol{y}(t)$ with interval $T_s=1/B$, where $B=N\Delta f$ is the signal bandwidth. The received signal after sampling can be expressed as
\begin{equation}
\begin{aligned}
\abovedisplayshortskip=2pt
\belowdisplayshortskip=2pt
\abovedisplayskip=2pt
\belowdisplayskip=2pt
&\boldsymbol{y}\left[ k \right] =\boldsymbol{y}\left( kT_s \right) =\sum_{l=1}^L{\boldsymbol{h}_ls\left( kT_s-\tau _{s,l} \right)e^{j2\pi f_{D,l}kT_s}} +\boldsymbol{n}\left( kT_s \right)
\\
&=\sum_{l=1}^L{\boldsymbol{a}\left( r_{B,l},\theta _l \right) \alpha _ls\left( kT_s-\tau _{s,l} \right)e^{j2\pi f_{D,l}kT_s}} +\boldsymbol{n}\left( kT_s \right).
\end{aligned}
\label{ReceivedSignalSampled}
\end{equation}

\section{Single-BS Environment Sensing and UE Localization without LoS Path} \label{signalprocessing}
\begin{figure}[!htbp]
\vspace{-12pt}
\centering
\includegraphics[width=0.48\textwidth]{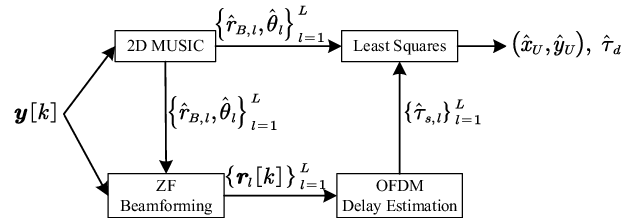}
\hfil
\vspace*{-6pt}
\caption{Signal processing flow of the proposed method for simultaneous environment sensing and UE localization.}
	\label{flowgraph}
\vspace*{-16pt}
\end{figure}

Fig. \ref{flowgraph} depicts the overall signal processing flow of the proposed method for environment sensing and UE localization. Initially, the two-dimensional multiple signal classification (2D-MUSIC) algorithm is utilized to estimate the positions $\{(r_{B,l},\theta_l)\}_{l=1}^{L}$ of the scatterers. Based on these position estimates, near-field zero-forcing (ZF) beamformers are constructed to isolate the signals from each path. Subsequently, OFDM radar algorithms are employed on the isolated signals to estimate the delays $\{\tau_{s,l}\}_{l=1}^L$. Finally, the estimated scatterer positions $\{(\hat{r}_{B,l},\hat{\theta}_l)\}_{l=1}^{L}$ and delays $\{\hat{\tau}_{s,l}\}_{l=1}^L$ are amalgamated to determine the UE position $(x_U,y_U)$ and clock difference $\tau_d$.

\vspace*{-8pt}
\subsection{Scatterer Localization} \label{localization}
The locations, i.e. the distance-AoA pairs of the scatterers $\{(r_{B,l},\theta_l)\}_{l=1}^{L}$ can be obtained from $\boldsymbol{y}\left[ k \right]$ by leveraging the near-field effects of the XL-ULA. To this end, the near-field 2D-MUSIC algorithm is employed. Let
\begin{equation}
\begin{aligned}
\begin{cases}
	\boldsymbol{S}\triangleq \left[ \boldsymbol{s}_1,\cdots ,\boldsymbol{s}_L \right] ^T\in \mathbb{C} ^{L\times K}\\
	\boldsymbol{A}\triangleq \left[ \boldsymbol{a}\left( r_{B,1},\theta _1 \right) ,\cdots ,\boldsymbol{a}\left( r_{B,L},\theta _L \right) \right] \in \mathbb{C} ^{M\times L}\\
	\boldsymbol{N}\triangleq \left[ \boldsymbol{n}\left[ 0 \right] ,\cdots ,\boldsymbol{n}\left[K-1\right] \right] \in \mathbb{C} ^{M\times K}\\
	\boldsymbol{Y}\triangleq \left[ \boldsymbol{y}\left[ 0 \right] ,\cdots ,\boldsymbol{y}\left[ K-1 \right] \right] \in \mathbb{C} ^{M\times K}\\
\end{cases},
\end{aligned}
\label{Matrixnotations}
\end{equation}
where $K=\lceil {(\varGamma T_O+\tau _{s,\max})}/{(N\Delta f)} \rceil $ is the number of samples, $\boldsymbol{s}_l\triangleq \left[ s_l\left[ 0 \right] ,\cdots ,s_l\left[ K-1 \right] \right] ^T\in \mathbb{C} ^{K\times 1}$, $s_l\left[ k \right] \triangleq \alpha _ls\left( kT_s-\tau_{s,l} \right)e^{j2\pi f_{D,l}kT_s}$
denotes the signal received from the $l$th path and $
\boldsymbol{n}\left[ k \right] \triangleq \boldsymbol{n}\left( kT_s \right) 
$ is the sampled noise. $\boldsymbol{Y}$ can be further expressed as
\begin{equation}
\begin{aligned}
\boldsymbol{Y}=\sum_{l=1}^L{\boldsymbol{a}\left( r_{B,l},\theta _l \right) \boldsymbol{s}_{l}^{T}}+\boldsymbol{N}=\boldsymbol{AS}+\boldsymbol{N}.
\end{aligned}
\label{Received matrix}
\end{equation}

The correlation matrix of the received signal is $\boldsymbol{R}_y=\boldsymbol{YY}^H$. Eigenvalue decomposition (EVD) is then performed on $\boldsymbol{R}_y$ as $\boldsymbol{R}_y=\boldsymbol{E}_s\boldsymbol{\varLambda}_s\boldsymbol{E}_{s}^{H}+\boldsymbol{E}_n\boldsymbol{\varLambda }_n\boldsymbol{E}_{n}^{H}$, where $\boldsymbol{\varLambda}_s$ and $\boldsymbol{\varLambda}_n$ denote the diagonal matrices consisting of $L$ large eigenvalues and $M-L$ small eigenvalues respectively, and the columns of $\boldsymbol{E}_s$ and $\boldsymbol{E}_n$ are the corresponding eigenvectors. The linear span of the column vectors in $\boldsymbol{E}_s$ and $\boldsymbol{E}_n$ are called the \emph{signal subspace} and \emph{noise subspace} respectively.
Then, by leveraging the fact that the noise subspace is orthogonal to the signal subspace, the column vectors of $\boldsymbol{E}_n$ are used to calculate the MUSIC spectrum,
\begin{equation}
\begin{aligned}
P_{\mathrm{MUSIC}}\left( r_B,\theta \right) &=\frac{1}{\boldsymbol{a}^H\left( r_B,\theta \right) \boldsymbol{E}_n\boldsymbol{E}_{n}^{H}\boldsymbol{a}\left( r_B,\theta \right)}
\\
&=\frac{1}{M-\boldsymbol{a}^H\left( r_B,\theta \right) \boldsymbol{E}_s\boldsymbol{E}_{s}^{H}\boldsymbol{a}\left( r_B,\theta \right)},
\end{aligned}
\label{MUSIC}
\end{equation}
where $\boldsymbol{a}\left( r_B,\theta \right)$ is the array response vector given in (\ref{ArrayVector}). The locations of the scatterers $\{(r_{B,l},\theta_l)\}_{l=1}^{L}$ can then be estimated by searching the spectrum. 

Note that the signals from different scatterers $\{\boldsymbol{s}_l\}_{l=1}^L$ are delayed and Doppler-shifted versions of the same transmitted OFDM signal. Therefore, unlike the case with single-tone signals, they are linearly independent as long as all the delay-Doppler pairs are different. However, when the delay-Doppler pairs of any two paths are very close, the signal sources $\{\boldsymbol{s}_l\}_{l=1}^{L}$ may become too correlated, degrading the MUSIC performance\cite{pillai1989performance}. Traditionally, this is solved by decorrelation techniques like spatial smoothing \cite{spatialsmoothing}, but these techniques require the array manifold $\boldsymbol{A}$ to be Vandermonde and are thus not directly applicable to near-field scenarios\cite{pannert2011spatial}. Moreover, there are high sidelobes in the range axis of the MUSIC spectrum, which might be falsely detected as scatterers, especially when the true peaks are low due to weak scatterers or high signal correlation. To solve these issues and improve accuracy, we propose a modified version of the sequential ZF MUSIC algorithm\cite{oh1993sequential}, named successive ZF 2D-MUSIC, as described in Algorithm \ref{Sequential ZF-MUSIC}.

\begin{algorithm}
    \caption{Successive ZF 2D-MUSIC}
    \begin{algorithmic}[1]
        \REQUIRE 
        $\boldsymbol{R}_y \in \mathbb{C}^{M \times M}$ 
        \ENSURE Set $\mathcal{S}$ containing distance-AoA pairs $(\hat{r}_{B,l},\hat{\theta }_l)$, number of paths $\hat{L}$ 
        \STATE {$\mathcal{S} = \varnothing$, $\boldsymbol{C}_1=[\,]$, $\boldsymbol{W}_1 = \boldsymbol{I}_M$, $l=1$, $\boldsymbol{R}_1=\boldsymbol{R}_y$}
        \REPEAT
            \STATE Obtain the signal subspace matrix $\boldsymbol{E}_{s,l}$ for $\boldsymbol{R}_l$\\
            \STATE Define $\boldsymbol{b}_l\left( r_B,\theta \right) \triangleq \boldsymbol{W}_l\boldsymbol{a}\left( r_B,\theta \right) /\left\| \boldsymbol{W}_l\boldsymbol{a}\left( r_B,\theta \right) \right\| $, calculate the 2D spectrum $P_l\left( r_B,\theta \right)$ for $\boldsymbol{R}_l$ as
            $$P_l\left( r_B,\theta \right) =\left\| \boldsymbol{E}_{s,l}^{H}\boldsymbol{b}_l\left( r_B,\theta \right) \right\| ^2$$\\
            \STATE Find the maximum value $P_l(\hat{r}_{B,l},\hat{\theta}_l)$ in $P_l\left( r_B,\theta \right)$
            \IF{$P_l(\hat{r}_{B,l},\hat{\theta}_l)>P_{th}$}
            \STATE $\mathcal{S} =\mathcal{S} \cup (\hat{r}_{B,l},\hat{\theta}_l)$\\
            \STATE $\boldsymbol{C}_{l+1}=[\boldsymbol{C}_l,\boldsymbol{a}(\hat{r}_{B,l},\hat{\theta}_l)]$\\
            \STATE $\boldsymbol{W}_{l+1}=\boldsymbol{I}_M-\boldsymbol{C}_{l+1}\left( \boldsymbol{C}_{l+1}^H\boldsymbol{C}_{l+1} \right) ^{-1}\boldsymbol{C}_{l+1}^H$\\     
            \STATE $\boldsymbol{R}_{l+1}=\boldsymbol{W}_{l+1}\boldsymbol{R}_y\boldsymbol{W}_{l+1}^H$
            \STATE $l=l+1$
            \ENDIF
        \UNTIL{$P_l(\hat{r}_{B,l},\hat{\theta_l }) \leq P_{th}$}
        \STATE $\hat{L}=l-1$
    \end{algorithmic}
\label{Sequential ZF-MUSIC}
\end{algorithm}

In Algorithm \ref{Sequential ZF-MUSIC}, a technique akin to successive interference cancellation (SIC) is employed, where scatterer locations are successively estimated from the strongest to the weakest. Following the estimation of each scatterer location, the signal from the corresponding path is suppressed using ZF beamforming. For example, after the first scatterer is localized at $(\hat{r}_{B,1},\hat{\theta}_1)$, the received signal from the corresponding path is zero-forced, yielding $\boldsymbol{Y}_2=\boldsymbol{W}_2\boldsymbol{Y}$, where $\boldsymbol{W}_2=\left( \boldsymbol{I}_M-\hat{\boldsymbol{a}}_1\left( \hat{\boldsymbol{a}}_{1}^{H}\hat{\boldsymbol{a}}_1 \right) ^{-1}\boldsymbol{a}_{1}^{H} \right)$ and $\hat{\boldsymbol{a}}_1=\boldsymbol{a}(\hat{r}_{B,1},\hat{\theta}_1)$. The correlation matrix of $\boldsymbol{Y}_2$ can then be calculated as $\boldsymbol{R}_2=\boldsymbol{Y}_2\boldsymbol{Y}_{2}^{H}=\boldsymbol{W}_2\boldsymbol{R}_y\boldsymbol{W}_{2}^{H}$. Then, we search for the maximum value in $\boldsymbol{R}_2$'s 2D-MUSIC spectrum defined in (\ref{MUSIC}). Note that to ensure orthogonality between the array vectors and the noise subspace, the array vectors used for 2D-MUSIC calculation are also zero-forced using $\boldsymbol{W}_2$ and then normalized, i.e., the 2D-MUSIC spectrum for $\boldsymbol{R}_2$ is calculated as follows,
\begin{equation}
\begin{aligned}
P_{\mathrm{MUSIC},2}\left( r_B,\theta \right) = \frac{1}{M\left(1-\left\| \boldsymbol{E}_{s,2}^{H}\boldsymbol{b}_{2}\left( r_B,\theta \right)  \right\| ^2\right)},
\end{aligned}
\label{ZF_MUSIC_part}
\end{equation}
where $\boldsymbol{b}_2\left( r_B,\theta \right) \triangleq \boldsymbol{W}_2\boldsymbol{a}\left( r_B,\theta \right) /\left\| \boldsymbol{W}_2\boldsymbol{a}\left( r_B,\theta \right) \right\|$. Since we only need to find the maximum value in (\ref{ZF_MUSIC_part}), to reduce computational complexity, $P_{2}\left( r_B,\theta \right)$ can be calculated as in step 4 of Algorithm \ref{Sequential ZF-MUSIC} by letting $l=2$, instead of calculating $P_{\mathrm{MUSIC},2}\left( r_B,\theta \right)$. After each iteration, the maximum value $P_l(\hat{r}_{B,l},\hat{\theta}_l)$ is compared with a threshold $P_{th}\in(0,1)$ to determine whether all the scatterers are found. When $(\hat{r}_{B,l},\hat{\theta}_l)$ corresponds to a true target, $P_l(\hat{r}_{B,l},\hat{\theta}_l)$ is close to $1$. In our simulations below, $P_{th}$ is empirically set to $0.5$.
\vspace{-8pt}
\subsection{Path Isolation}
To avoid complex joint estimation, after locating the scatterers, the reflected signal from each scatterer is isolated using receiver-side beamforming. The isolated signal from the $l$th path can be written as
\begin{equation}
\begin{aligned}
r_l\left[ k \right] &=\boldsymbol{f}_{l}^{H}\boldsymbol{y}\left[ k \right] 
\\
&=\sum_{l^{\prime}=1}^L{\boldsymbol{f}_{l}^{H}\boldsymbol{h}_{l^{\prime}}s\left( kT_s-\tau _{s,l^{\prime}} \right)e^{j2\pi f_{D,l^{\prime}}kT_s}}+\boldsymbol{f}_{l}^{H}\boldsymbol{n}\left[ k \right].
\end{aligned}
\label{beamforming}
\end{equation}
To ensure perfect isolation between the separated signals, $\{\boldsymbol{f}_l\}_{l=1}^{L}$ should be designed so that $\boldsymbol{f}_{l}^{H}\boldsymbol{h}_{l^{\prime}}=0,\,\forall l^{\prime}\ne l$,
then the separated signal in (\ref{beamforming}) can be written as
\begin{equation}
\begin{aligned}
r_l\left[ k \right] =\boldsymbol{f}_{l}^{H}\boldsymbol{h}_{l}s\left( kT_s-\tau _{s,l} \right)e^{j2\pi f_{D,l}kT_s} +\boldsymbol{f}_{l}^{H}\boldsymbol{n}\left[ k \right].
\end{aligned}
\label{beamforming2}
\end{equation}
Next, we aim to maximize the signal-to-noise ratio (SNR) of (\ref{beamforming2}) under the ZF condition,
\begin{equation}
\begin{aligned}
&\underset{\left\{ \boldsymbol{f}_l \right\} _{l=1}^{L}}{\mathrm{maximize}}\frac{P\left| \boldsymbol{f}_{l}^{H}\boldsymbol{h}_l \right|^2}{\sigma ^2\left\| \boldsymbol{f}_{l} \right\| ^2}
\\
&\mathrm{s.t.}  \,\,\boldsymbol{f}_{l}^{H}\boldsymbol{h}_{l^{\prime}}=0, \,\forall l^{\prime}\ne l; \left\| \boldsymbol{f}_{l} \right\|=1.
\end{aligned}
\label{ZF_opti}
\end{equation}
Let $\boldsymbol{H}_{l}\triangleq[\boldsymbol{h}_1,\ldots,\boldsymbol{h}_{l-1},$ $\boldsymbol{h}_{l+1},\ldots,\boldsymbol{h}_{L}]$ and denote by 
$\boldsymbol{Q}_l\triangleq \boldsymbol{I}_M-\boldsymbol{H}_l\left( \boldsymbol{H}_{l}^{H}\boldsymbol{H}_l \right) ^{-1}\boldsymbol{H}_{l}^{H}$ the projection matrix into the null space of $\boldsymbol{H}_{l}^{H}$, the optimal ZF beamforming is $\boldsymbol{f}_l=e^{j\phi}{{\boldsymbol{Q}_l\boldsymbol{h}_l}/{\left\| \boldsymbol{Q}_l\boldsymbol{h}_l \right\|}}$, where $\phi$ is an arbitrary phase. Since $\boldsymbol{h}_l=\alpha _l\boldsymbol{a}\left( r_{B,l},\theta _l \right) $, the above ZF beamforming vectors can be estimated using the parameters $\{(\hat{r}_{B,l},\hat{\theta }_l)\}_{l=1}^{\hat{L}}$ obtained in \ref{localization} as $\hat{\boldsymbol{f}}_l=\hat{\boldsymbol{Q}}_l\hat{\boldsymbol{a}}_l/{\| \hat{\boldsymbol{Q}}_l\hat{\boldsymbol{a}}_l\| }$, where $\hat{\boldsymbol{a}}_l=\boldsymbol{a}\left( \hat{r}_{B,l},\hat{\theta}_l \right)$, $\hat{\boldsymbol{Q}}_l=\boldsymbol{I}_M-\boldsymbol{A}_l\left( \boldsymbol{A}_{l}^{H}\boldsymbol{A}_l \right) ^{-1}\boldsymbol{A}_{l}^{H}$ and $
\boldsymbol{A}_l=[\hat{\boldsymbol{a}}_1,...,\hat{\boldsymbol{a}}_{l-1},\hat{\boldsymbol{a}}_{l+1},...,\hat{\boldsymbol{a}}_{\hat{L}}]$, $l=1,\ldots,\hat{L}$.
\subsection{Delay Estimation}
As shown in (\ref{beamforming2}), after ZF beamforming, each separated signal has only one delay $\tau_{s,l}$ and one Doppler shift $f_{D,l}$, which can be estimated using OFDM radar methods\cite{2014ofdmradar}. Under the assumption that the maximum delay $\tau _{s,\max}={\max}\left\{ \tau_{s,l} \right\}$ does not exceed the CP duration $T_{CP}$, and that the maximum Doppler shift $f _{D,\max}={\max}\left\{ f_{D,l} \right\}$ is smaller than $\Delta f/10$, then after CP removal, the received signal can be rearranged into $N\times \varGamma $ receive matrices $\left\{ \boldsymbol{F}_{\mathrm{Rx},l} \right\} _{l=1}^{\hat{L}}$ using the fast Fourier transform (FFT)\cite{2014ofdmradar},
\begin{equation}
\begin{aligned}
\left( \boldsymbol{F}_{\mathrm{Rx},l} \right) _{n,\gamma}\!\!&=\underbrace{\!\hat{\boldsymbol{f}}_{l}^{H}\boldsymbol{h}_lb_{n,\gamma}e^{j2\pi (\gamma f_{D,l}T_O-n\Delta f\tau _{s,l})}}_{\mathrm{desired\,signal}}
\\
&+\underbrace{\sum_{l^{\prime}\ne l}^L{\!\hat{\boldsymbol{f}}_{l}^{H}\boldsymbol{h}_{l^{\prime}}b_{n,\gamma}e^{j2\pi (\gamma f_{D,l^{\prime}}T_O-n\Delta f\tau _{s,l^{\prime}})}}}_{\mathrm{inter-path\,interference}}\!+\left( \!\boldsymbol{Z}_l \right) _{n,\gamma},
\end{aligned}
\label{F_Rx_l}
\end{equation}
where $\left( \cdot \right) _{n,\gamma}$ denotes the $(n,\gamma)$th element of a matrix, $\left\{ \mathbf{Z}_{l} \right\} _{l=1}^{L}$ are the noise matrices. The first term in \eqref{F_Rx_l} corresponds to the signal from the $l$th path and the second term is inter-path interference caused by imperfect estimation of the scatterer locations $\{(\hat{r}_{B,l},\hat{\theta }_l)\}_{l=1}^{\hat{L}}$. The influence of transmitted data symbols can be removed by element-wise division,
\begin{equation}
\begin{aligned}
\left( \boldsymbol{F}_l \right) _{n,\gamma}&=\!\hat{\boldsymbol{f}}_{l}^{H}\boldsymbol{h}_le^{j2\pi (\gamma f_{D,l}T_O-n\Delta f\tau _{s,l})}
\\
&+\sum_{l^{\prime}\ne l}^L{\!\hat{\boldsymbol{f}}_{l}^{H}\boldsymbol{h}_{l^{\prime}}e^{j2\pi (\gamma f_{D,l^{\prime}}T_O-n\Delta f\tau _{s,l^{\prime}})}}+(\tilde{\boldsymbol{Z}}_l)_{n,\gamma},
\end{aligned}
\label{F_l}
\end{equation}
where $( \tilde{\boldsymbol{Z}}_l) _{n,\gamma}={\left( \!\boldsymbol{Z}_l \right) _{n,\gamma}/{b_{n,\gamma}}}$. The periodogram method \cite{zhang2023integrated} can then be used to accurately estimate the delay $\tau_{s,l}$. Note that since we do not need to estimate the Doppler here, the periodogram can be calculated as
\begin{equation}
\begin{aligned}
\mathrm{Per}_l\left[ k \right] =\frac{1}{N\varGamma}\sum_{\gamma =0}^{\varGamma-1}{\left| \sum_{n=0}^{N-1}{\left( \boldsymbol{F}_l \right) _{n,\gamma}w\left[ n \right] e^{j2\pi \frac{kn}{N_{\mathrm{Per}}}}} \right|^2},
\end{aligned}
\label{Per}
\end{equation}
where $w[n]$ is the window function used to suppress the sidelobes. (\ref{Per}) can be efficiently calculated using inverse fast Fourier transform (IFFT) of length $N_{\mathrm{Per}}$.
Note that the second term in \eqref{F_l} is relatively small when the parameters $\{(r_{B,l},\theta_l)\}_{l=1}^{L}$ are estimated accurately. Then, by finding the maximum point $\hat{k}_l$ in $\mathrm{Per}_l\left[ k \right]$, the sum of the clock difference $\tau_d$ and the propagation delay $\tau_l$ for the $l$th path is estimated as
\begin{equation}
\begin{aligned}
\hat{\tau}_{s,l} =\frac{\hat{k}_l}{N_{\mathrm{Per}}\Delta f}.
\end{aligned}
\label{delay_es}
\vspace*{-9pt}
\end{equation}
\subsection{UE Localization} \label{ueloc}
Since $\tau _l=(r_{B,l}+r_{U,l})/c$, ${r}_{U,l}$ can be expressed as
\begin{equation}
\begin{aligned}
{r}_{U,l}=c{\tau}_{l}-{r}_{B,l}=c({\tau}_{s,l}-\tau_d)-{r}_{B,l}.
\end{aligned}
\label{r_ul_es}
\end{equation}
Now that the locations of the scatterers $\{(\hat{r}_{B,l},\hat{\theta}_l)\}_{l=1}^{\hat{L}}$ as well as the sums of the clock difference and the propagation delays $\left\{ \hat{\tau}_{s,l} \right\} _{l=1}^{\hat{L}}$ have been estimated, the location of the UE $ \left( x_U,y_U \right)$ and the clock difference $\tau_d$ can be estimated by solving the following equations,
\begin{equation}
\begin{aligned}
\left\{ \begin{array}{c}
	(\hat{x}_l-x_U)^2+(\hat{y}_l-y_U)^2=\hat{r}_{U,l}^{2}\\
	\hat{r}_{U,l}=c(\hat{\tau}_{s,l}-\tau_d)-\hat{r}_{B,l}\\
\end{array}, \forall l=1,\ldots,\hat{L} \right.,
\end{aligned}
\label{pos_ue_es}
\end{equation}
where $\hat{x}_l=\hat{r}_{B,l}\mathrm{cos}(\hat{\theta }_l) $ and $\hat{y}_l=\hat{r}_{B,l}\mathrm{sin}(\hat{\theta }_l)$. Assuming $\hat{L}\ge4$, denote by $\hat{q}_l=c\hat{\tau}_{s,l}-\hat{r}_{B,l}$, then the above equation system can be converted to a linear equation $\boldsymbol{D}\boldsymbol{x}_u=\boldsymbol{p}$, where $\boldsymbol{x}_u= \left[ x_U,y_U,\tau_d \right]^T$, $\boldsymbol{D}$ and $\boldsymbol{p}$ can be expressed as
\begin{equation}
\begin{aligned}
\boldsymbol{D}&=\left[ \begin{matrix}
	2\left( \hat{x}_1-\hat{x}_2 \right)&		2(\hat{y}_1-\hat{y}_2)&		2(\hat{q}_2-\hat{q}_1)c\\
	\vdots&		\vdots&		\vdots\\
	2\left( \hat{x}_1-\hat{x}_L \right)&		2(\hat{y}_1-\hat{y}_L)&		2(\hat{q}_L-\hat{q}_1)c\\
\end{matrix} \right] \in \mathbb{R} ^{(\hat{L}-1)\times 3}
\\
\boldsymbol{p}&=\left[ \begin{array}{c}
	\hat{x}_{1}^{2}-\hat{x}_{2}^{2}+\hat{y}_{1}^{2}-\hat{y}_{2}^{2}+\hat{q}_{2}^{2}-\hat{q}_{1}^{2}\\
	\vdots\\
	\hat{x}_{1}^{2}-\hat{x}_{L}^{2}+\hat{y}_{1}^{2}-\hat{y}_{L}^{2}+\hat{q}_{L}^{2}-\hat{q}_{1}^{2}\\
\end{array} \right] \in \mathbb{R} ^{\hat{L}-1}.
\end{aligned}
\label{D and p}
\end{equation}
The above problem can be solved using least squares when $\hat{L}\ge4$,
\begin{equation}
\begin{aligned}
\hat{\boldsymbol{x}}_u=\left( \boldsymbol{D}^T\boldsymbol{D} \right) ^{-1}\boldsymbol{D}^T\boldsymbol{p}.
\end{aligned}
\label{LS_solve}
\end{equation}

\section{Simulation Results}\label{Sim}
Simulations are conducted to evaluate the performance of the proposed scheme. The transmitted signal has a bandwidth of $B = 400 \,\mathrm{MHz}$,  a CP length of $T_{CP}=256/B=0.64\,\mathrm{\mu s}$, number of subcarriers $N = 1024$, and number of OFDM symbols $\varGamma=100$. The carrier frequency is $f_c = 28 \,\mathrm{GHz}$. The number of BS antennas is $M = 256$, and the antenna spacing is $d=\lambda/2$, with the $m$th antenna element at $(0,md)$. The UE is located at $(15\sqrt{3},15)\,\mathrm{m}$ with velocity $(0,10) \,\mathrm{m/s}$. and the number of scatterers is $L=5$. The received SNR varies from $-10\,\mathrm{dB}$ to $20\,\mathrm{dB}$. The following results are averaged over 100 Monte Carlo simulations.

\vspace{-9pt}
\begin{table}[htbp]
	\centering  
	\caption{Locations of close and far-apart scatterers}  
	\label{tb1}  
	\begin{tabular}{c|c|c|c|c|c}  
		\hline  
		$l$&1&2&3&4&5 \\  
		\hline
		$R_{B,l}/\mathrm{m}$ (C)&19.9&20.9&20.3&2.1&7.8 \\
        \hline
  	$\theta_l/^{\circ}$ (C)&-18&-19&-21&-24&14 \\
        \hline
   	$R_{B,l}/\mathrm{m}$ (F)&26.6&5.7&23.0&17.8&15.4 \\
        \hline
    	$\theta_l/^{\circ}$ (F)&11&-23&57&-16&-6 \\
		\hline
	\end{tabular}
\end{table}
\vspace{-9pt}
\begin{figure}[!htbp]
  \vspace{-20pt}
  \captionsetup[subfloat]{labelfont=footnotesize,textfont=footnotesize,oneside,margin={0.55cm,0cm}}
  \subfloat[UE position $(\hat{x}_U,\hat{y}_U)$]{\includegraphics[width=0.233\textwidth]{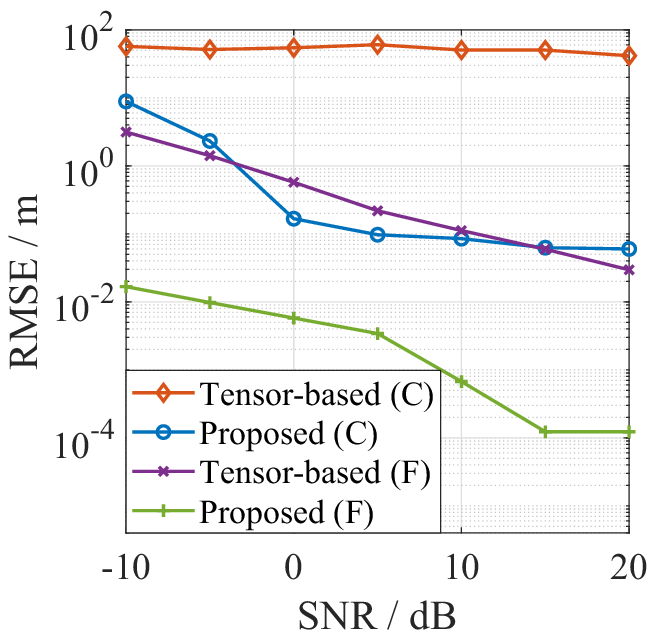}}
  \subfloat[clock difference $\hat{\tau}_d$]{\includegraphics[width=0.247\textwidth]{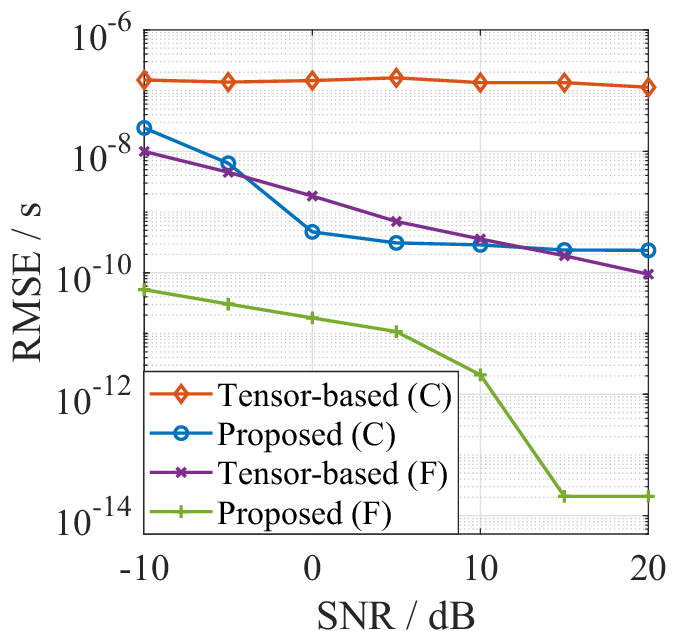}}
  \caption{RMSE of the estimated parameters for close (C) and far-apart (F) scatterers.}
  \label{fig_RMSE}
  \vspace{-6pt}
\end{figure}

Fig. \ref{fig_RMSE} shows the resulting root mean square error (RMSE) of the estimated UE position $(\hat{x}_U,\hat{y}_U)$ and clock difference $\hat{t}_d$ for close and far-apart scatterers whose locations are shown in Table \ref{tb1}. For the tensor-based method, the closed-form CP decomposition \cite{9048762} is used for its low complexity and robustness to low Doppler. 
It is observed that the proposed method outperforms the tensor-based method for both cases. Specifically, when the scatterers are far apart, our method achieves significantly better accuracy than the tensor-based method. When the scatterers are close, the tensor-based method fails to separate the different paths and therefore performs poorly, while our approach still works, demonstrating superior robustness to closely spaced scatterers thanks to the super-resolution capability provided by 2D-MUSIC.


The complexity for the main steps in our method is as follows (assuming $L\ll M$). Algorithm \ref{Sequential ZF-MUSIC}: $\mathcal{O}(( \varGamma N+N_g) M^2+LM^3)$, where $N_g$ is the number of grids searched; path isolation: $\mathcal{O} \left( M^2L^2+L^4+LMN\varGamma \right) $; delay estimation: $\mathcal{O} \left( L\varGamma N_{\mathrm{Per}}\log \left( N_{\mathrm{Per}} \right) \right)$. Since $N_g$ and $\varGamma N$ are significantly larger than other variables, the complexity is dominated by $\mathcal{O}(( \varGamma N+N_g) M^2)$. In our simulation, the average run time for the tensor-based method and our method is $15.05\mathrm{s}$ and $7.49\mathrm{s}$, respectively.

\section{Conclusion}
This letter introduces a novel method for achieving simultaneous environment sensing and UE localization in the absence of LoS path using a single BS. Leveraging the near-field effects of XL-arrays, the positions of scatterers are estimated, which are subsequently utilized for NLoS UE localization. Simulation results demonstrate that the proposed method achieves superior accuracy and robustness compared with the existing tensor-based method.
\vspace{-6pt}
\bibliographystyle{IEEEtran}
\bibliography{IEEEabrv,reference}

\begin{thebibliography}{10}
\providecommand{\url}[1]{#1}
\csname url@samestyle\endcsname
\providecommand{\newblock}{\relax}
\providecommand{\bibinfo}[2]{#2}
\providecommand{\BIBentrySTDinterwordspacing}{\spaceskip=0pt\relax}
\providecommand{\BIBentryALTinterwordstretchfactor}{4}
\providecommand{\BIBentryALTinterwordspacing}{\spaceskip=\fontdimen2\font plus
\BIBentryALTinterwordstretchfactor\fontdimen3\font minus \fontdimen4\font\relax}
\providecommand{\BIBforeignlanguage}[2]{{%
\expandafter\ifx\csname l@#1\endcsname\relax
\typeout{** WARNING: IEEEtran.bst: No hyphenation pattern has been}%
\typeout{** loaded for the language `#1'. Using the pattern for}%
\typeout{** the default language instead.}%
\else
\language=\csname l@#1\endcsname
\fi
#2}}
\providecommand{\BIBdecl}{\relax}
\BIBdecl

\bibitem{xiao2022overview}
Z.~Xiao and Y.~Zeng, ``An overview on integrated localization and communication towards {6G},'' \emph{Sci. China Inf. Sci.}, vol.~65, no.~3, p. 131301, Dec. 2022.

\bibitem{localization_application2}
F.~Wen, H.~Wymeersch, B.~Peng, W.~P. Tay, H.~C. So, and D.~Yang, ``A survey on {5G} massive {MIMO} localization,'' \emph{Digit. Signal Process.}, vol.~94, pp. 21--28, Nov. 2019.

\bibitem{malm2018user}
N.~Malm, L.~Zhou, E.~Menta, K.~Ruttik, R.~J{\"a}ntti, O.~Tirkkonen, M.~Costa, and K.~Lepp{\"a}nen, ``User localization enabled ultra-dense network testbed,'' in \emph{Proc. IEEE 5G World Forum (5GWF)}.\hskip 1em plus 0.5em minus 0.4em\relax IEEE, Jul. 2018, pp. 405--409.

\bibitem{zeng2021toward}
Y.~Zeng and X.~Xu, ``Toward environment-aware {6G} communications via channel knowledge map,'' \emph{IEEE Wireless Commun.}, vol.~28, no.~3, pp. 84--91, Mar. 2021.

\bibitem{he2012modeling}
J.~He, Y.~Geng, and K.~Pahlavan, ``Modeling indoor {TOA} ranging error for body mounted sensors,'' in \emph{Proc. 23rd Int. Symp. Pers., Indoor Mobile Radio Commun. (PIMRC)}.\hskip 1em plus 0.5em minus 0.4em\relax IEEE, Sep. 2012, pp. 682--686.

\bibitem{shen2010fundamental}
Y.~Shen and M.~Z. Win, ``Fundamental limits of wideband localization—{Part I: A general framework},'' \emph{{IEEE} Trans. Inf. Theory}, vol.~56, no.~10, pp. 4956--4980, Sep. 2010.

\bibitem{al2002ml}
S.~Al-Jazzar and J.~Caffery, ``{ML} and {Bayesian} {TOA} location estimators for {NLOS} environments,'' in \emph{Proc. IEEE Vehicular Technology Conf. (VTC)}, vol.~2.\hskip 1em plus 0.5em minus 0.4em\relax IEEE, Sep. 2002, pp. 1178--1181.

\bibitem{liu2003toa}
L.~Liu, P.~Deng, and P.~Fan, ``A {TOA} reconstruction method based on ring of scatterers model,'' in \emph{Proc. Int. Conf. Parallel Distrib. Comput., Appl. Technol.}\hskip 1em plus 0.5em minus 0.4em\relax IEEE, Aug. 2003, pp. 375--377.

\bibitem{cao2024unified}
M.~Cao, H.~Zhang, B.~Di, and H.~Zhang, ``Unified near-field and far-field localization with holographic {MIMO},'' \emph{arXiv preprint arXiv:2401.06334}, 2024.

\bibitem{wang2024cramer}
H.~Wang, Z.~Xiao, and Y.~Zeng, ``Cram{\'e}r-rao bounds for near-field sensing with extremely large-scale {MIMO},'' \emph{{IEEE} Trans. Signal Process.}, Jan. 2024.

\bibitem{tensormethod}
I.~Podkurkov, G.~Seidl, L.~Khamidullina, A.~Nadeev, and M.~Haardt, ``Tensor-based near-field localization using massive antenna arrays,'' \emph{{IEEE} Trans. Signal Process.}, vol.~69, pp. 5830--5845, Aug. 2021.

\bibitem{nearfield}
H.~Lu and Y.~Zeng, ``Communicating with extremely large-scale array/surface: Unified modeling and performance analysis,'' \emph{{IEEE} Trans. Wireless Commun.}, vol.~21, no.~6, pp. 4039--4053, Nov. 2022.

\bibitem{pillai1989performance}
S.~U. Pillai and B.~H. Kwon, ``Performance analysis of {MUSIC}-type high resolution estimators for direction finding in correlated and coherent scenes,'' \emph{{IEEE} Trans. Acoust., Speech, Signal Process.}, vol.~37, no.~8, pp. 1176--1189, Aug. 1989.

\bibitem{spatialsmoothing}
T.-J. Shan, M.~Wax, and T.~Kailath, ``On spatial smoothing for direction-of-arrival estimation of coherent signals,'' \emph{{IEEE} Trans. Acoust., Speech, Signal Process.}, vol.~33, no.~4, pp. 806--811, Aug. 1985.

\bibitem{pannert2011spatial}
W.~Pannert, ``Spatial smoothing for localized correlated sources—{Its} effect on different localization methods in the nearfield,'' \emph{Appl. Acoust.}, vol.~72, no.~11, pp. 873--883, Nov. 2011.

\bibitem{oh1993sequential}
S.~K. Oh and C.~K. Un, ``A sequential estimation approach for performance improvement of eigenstructure-based methods in array processing,'' \emph{{IEEE} Trans. Signal Process.}, vol.~41, no.~1, p. 457, Jan. 1993.

\bibitem{2014ofdmradar}
K.~M. Braun, ``{OFDM} radar algorithms in mobile communication networks,'' Ph.D. dissertation, Karlsruhe, Karlsruher Institut f{\"u}r Technologie, Germany, 2014.

\bibitem{zhang2023integrated}
C.~Zhang, Z.~Zhou, H.~Wang, and Y.~Zeng, ``Integrated super-resolution sensing and communication with {5G} {NR} waveform: Signal processing with uneven cps and experiments,'' in \emph{Proc. 21st Int. Symp. Model. Optim. Mobile, Ad Hoc, Wireless Netw. (WiOpt)}.\hskip 1em plus 0.5em minus 0.4em\relax IEEE, Aug. 2023, pp. 681--688.

\bibitem{9048762}
Y.~Lin, S.~Jin, M.~Matthaiou, and X.~You, ``Structured tensor decomposition-based channel estimation for wideband millimeter wave {MIMO},'' in \emph{Proc. 53rd Asilomar Conf. Signals, Syst., Comput.}, Nov. 2019, pp. 421--426.

\end{thebibliography}

\end{document}